\documentclass[twocolumn,twoside,slac]{revtex4}
\usepackage{graphicx}
\usepackage{fancyhdr}
\begin{document}

\title{Simulation in ALICE}

\author{F. Carminati, A. Morsch \\ on behalf of the ALICE Offline Project}
\affiliation{CERN, 1211 Geneva 23, Switzerland}

\begin{abstract}
ALICE, the experiment dedicated to the study of heavy ion collisions at the LHC, 
uses an object-oriented framework for simulation, reconstruction and analysis (AliRoot) based on ROOT.  
Here, we describe the general ALICE simulation strategy and those components of the framework related 
to simulation. 
Two main requirements have driven the development of the simulation components. 
First, the possibility to run different transport codes with the same user code for geometry 
and detector response has led to the development of the Virtual Monte Carlo concept. 
Second, simulation has to provide tools to efficiently study events ranging from low-multiplicity 
pp collisions to Pb-Pb collisions with up to 80000 primary particles per event. 
This has led to the development of a variety of collaborating generator classes and specific classes 
for event merging.

\end{abstract}
\maketitle

\thispagestyle{fancy}

\begin{figure*}[hbt]
\centering
\includegraphics[width=140mm]{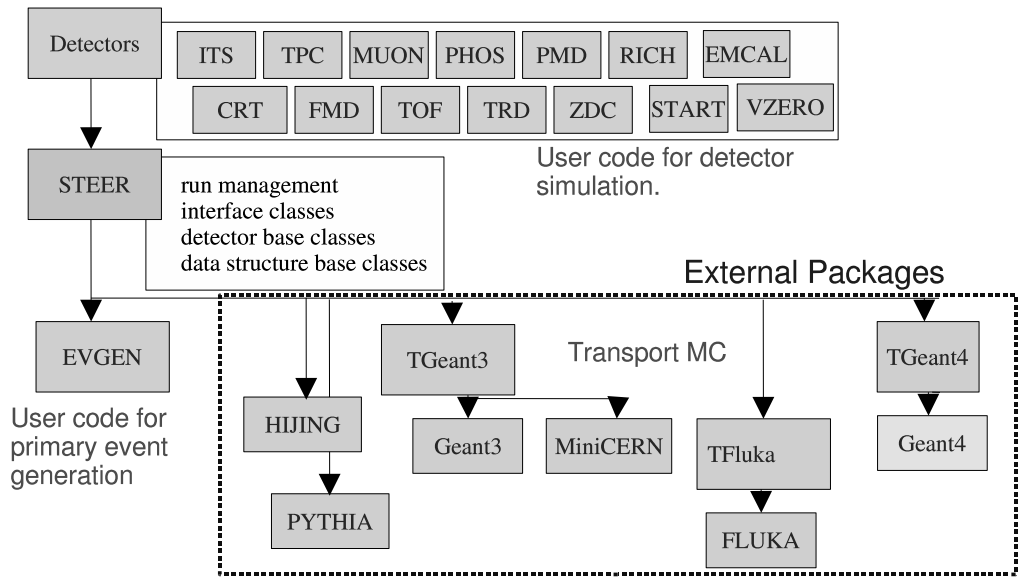}
\caption{Component view of the AliRoot simulation framework.}
\label{fig:aliroot}
\end{figure*}

\section{Introduction}
ALICE, is the experiment dedicated to the study of heavy ion collisions at the LHC.
It is a multipurpose detector with excellent tracking and secondary vertex capabilities, 
electron and muon detection and a high resolution $\gamma$-spectrometer. 
Although smaller in size as compared to the large LHC detectors ATLAS and CMS, 
ALICE is of similar complexity.
Instead of a complete description of the detector we mention here exemplarily that 
close to the vertex six layers of silicon trackers (pixel-, drift-, and strip-detectors) are
used mainly for secondary vertex reconstruction. 
The main tracking device is a very large Time Projection Chamber (TPC) covering approximatively 
two units of pseudo-rapidity. 
Moreover, ALICE uses almost all known particle identification techniques including 
Cerenkov and Transition Radiation detectors whose detailed response simulation 
is a demanding task.

Heavy ion collisions produce a very large number of particles in the final state. 
Current predictions range from 1400-8000 charged particles in the central unit of 
rapidity. 
This is a challenge for the reconstruction and analysis algorithms. 
The development of these algorithms requires a predictive and precise simulation. 
Particles have to be transported in the detectors and structural elements. 
The latter produce secondary particles which can increase significantly the detector 
occupancy and their correct prediction is important. 
For those particles passing through sensitive 
detector areas a detailed detector response simulation is necessary.

Transport and detector response simulation is only one part of the simulation task. 
Of comparable complexity and diversity is the primary event simulation. 
It comprises both the heavy ion physics specific soft physics and the hard probes like heavy 
flavor and jet production. 
In addition, ALICE has to simulate all possible collision systems ranging from pp, p-A, 
intermediate mass A--A, to  Pb--Pb as well as collision geometries ranging from central to peripheral collisions.

Before we describe the different components of the simulation framework
we outline in the following section the ALICE simulation strategy as it has been 
developed on the basis of the above mentioned challenges and requirements.  
 
\section{ALICE simulation strategy}

In order to cope with the challenges described in the introduction, the ALICE Offline Project has 
developed a coherent simulation framework as an integral part of the AliRoot \cite{aliroot} object oriented (C++) 
framework for simulation, reconstruction and analysis based on ROOT \cite{root}. 
It comprises primary event (physics) simulation, particle transport, detailed detector response simulation and 
fast simulation. 
Its main components are the Virtual MC, the detector classes containing the user code, 
a set of collaborating classes for primary particle simulation, and base classes for fast simulation.

Since the complex description of the detector geometries and of the detector responses has to be accomplished 
by a relatively small community of physicists it is essential to provide a stable framework which does
not require rewriting of the user code in the case that one of its underlying components changes. 
The most important of these underlying components is the transport MC. 
However, in traditional simulation environments the user code for geometry description, 
and detector response simulation (hit generation) depends entirely on the transport MC. 
ALICE uses currently Geant3 \cite{geant3} in production, FLUKA \cite{fluka} 
and Geant4 \cite{geant4} are both options for simulations in the near future. 
Other alternatives might come up after LHC start-up.
For this reason the ALICE Offline Project has developed the concept of the Virtual MC which 
provides an insulation layer between the user code and the underlying MC.

As already outlined in the introduction also primary event generation is a complex task.
It requires the generation of uncorrelated underlying events, correlation between particles and  
a multitude of so called hard probes as well as any possible mixture between these components. 
Generators come as external generators packages like Pythia and HIJING as well as simple 
user written generators based on parameterisations or so called {\it afterburners} which introduce
particle correlations in existing events. 
The simulation framework provides a set of collaborating base classes as building blocks 
for this variety of generators. 
These will be described in section \ref{sec:generators}. 

The size of events produced in heavy ion collisions and the variety of physics signals and collision 
systems which have to be studied require enormous amounts of computing time and data storage. 
The framework has to help to reduce the required resources as much as possible. 
One such tool is known as merging or embedding and will be described in the following section.

Another tool is fast simulation, which is needed for high statistics studies of signals and backgrounds 
for which the detector acceptances, reconstruction efficiencies, and resolutions have been 
already determined and parameterised using detailed simulations. 
The framework provides a set of base classes providing a uniform interface and facilitating the
realisation of fast simulation applications.

Last not least it should be mentioned that the ALICE simulation framework provides a rigorous approach 
concerning MC truth using the class {\it AliStack} developed for this purpose.
The history of each particle resulting into a detector hit and all primary particles are recorded. 
The user has the possibility to write additional information if needed.

\section{Simulation in the AliRoot framework}

\subsection{Components}

Fig. \ref{fig:aliroot} shows a component view of the AliRoot simulation framework. 
The central module, STEER, provides the interface classes for the detector description and 
data structures. 
It also provides the run steering and particle stack implementations required by the Virtual MC. 
The run steering communicates with the event generator module via the interface 
class {\it AliGenerator}, with the detector specific modules via {\it AliDetector},
and with the transport MC via the Virtual MC main interface class {\it TVirtualMC}.
The user code is structured according to the different sub-detectors.
No dependencies between these detector modules are allowed. 
The detector modules contain their specific implementations of 
{\it AliDetector} and data structures like {\it AliHit} and {\it AliDigit}.

Generator packages like HIJING and the {\it TVirtualMC} implementations {\it TGeant3}, {\it TGeant4} and 
{\it TFluka} appear as external modules. 
They are not needed for compilation and linking. The choice for their
usage can be made by the user during runtime.

\subsection{Simulated data}

The AliRoot simulation framework generates data at different stages of the simulation process. 
First, there are the so-called {\it hits} that represent the precise 
information about a particle passing a detector obtained from the transport MC, {\it i.e.} 
in most cases energy deposition and position. 
These {\it hits} are then transformed
into the signal produced by the detector, {\it summable digits} that
correspond to the raw data before addition of noise and threshold subtraction.  
The introduction of {\it summable digits} was necessary in order to  
realise the event merging strategy in which a underlying signal free event is 
combined with a signal event before completing the digitization process producing
the {\it digits} which contain the information of raw data. 
The importance of this procedure for the simulation of heavy ion collisions
lies in the fact that one underlying event can be used for several signal events 
thus reducing computation time and data storage space.

\begin{figure*}[hbt]
\centering
\includegraphics[width=140mm]{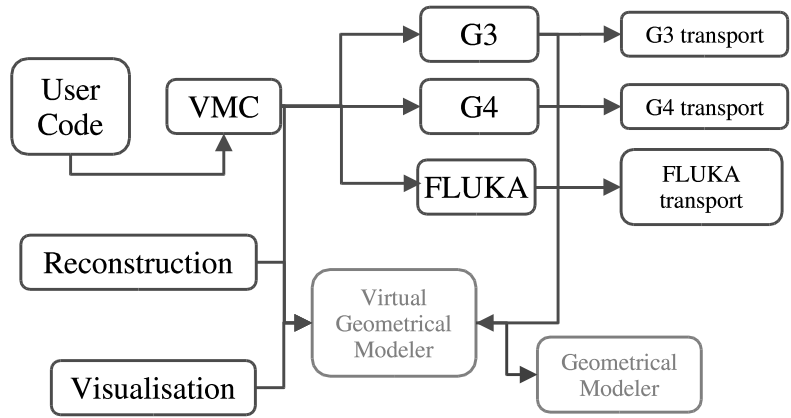}
\caption{Illustration of the Virtual MC concept.}
\label{fig:vmc}
\end{figure*}

\section{The Virtual Monte Carlo}
The VirtualMC interface has been developed by the ALICE Offline project in order to make the 
user code for detector simulation independent from the underlying transport MC. A detailed 
description can be found in Ref. \cite{vmc}.

An illustration of the VirtualMC concept is shown in Fig. \ref{fig:vmc}. 
A development that is ongoing (see Ref. \cite{tgeo}) is to separate the physics and transport part of the 
VirtualMC from the geometry modeler. 
This has the advantage that the geometry can be used independently from transport for reconstruction and visualisation. 
Moreover, the implementation of the geometry modeler can be optimised independently from the transport
MC. 

Currently ALICE uses {\it TGeant3}, the Geant3 implementation of {\it TVirtualMC} in production. 
{\it TGeant4} is used for simulations related to Geant4 physics validations. 
The {\it TFluka} implementation is almost completed and an intensive testing phase will follow before the 
first release.
Currently it uses FLUGG \cite{flugg} for geometry navigation. 
This is expected to be be replaced by the ROOT geometry modeler.

\section{Event generators}
\label{sec:generators}
\subsection{Simulation of heavy ion collisions}

The simulations of physical processes are confronted with several problems:

\begin {itemize}
\item Existing generators give different answers on parameters
  such as expected multiplicities, transverse momentum dependence and rapidity
  dependence at LHC energies.
\item Most of the physics signals, like Hyperon decays, high transverse momentum
phenomena, open charm and beauty, quarkonia etc., are not properly
reproduced by the existing generators.
\item Simulation of small cross sections would demand prohibitively
long runs to simulate a number of events that is commensurable with
the expected number of detected events in the experiment.
\item The existing generators do not provide for event topologies like momentum
correlations, azimuthal flow etc.
\end {itemize}

To allow nevertheless efficient simulations ALICE has adopted a
framework that allows for a number of options:

\begin{itemize}
\item{}
The simulation framework provides an interface to external generators, like
HIJING and DPMJET.

\item{} A parameterised ``signal free'' underlying event with multiplicity as
a parameter is provided.
\item{} Rare signals can be generated using the interface to external
generators like PYTHIA or simple parameterisations of transverse momentum and
rapidity spectra defined in function libraries.
\item{} The framework provides a tool to assemble events from different signal
generators (event cocktails).
\item{} The framework provides tools to combine underlying events and signal
events on the primary particle level (cocktail) and on the digit level
(merging).
\item{} {\it After-Burners} are used to introduce particle correlations in a
controlled way.
\end{itemize}

The main parts of the implementation of this strategy are described below.

\begin{figure}[htb]
\centering
\includegraphics[width=65mm]{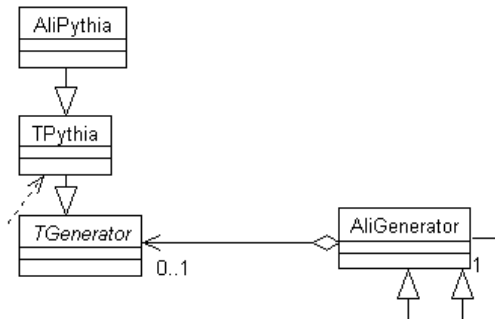}
\caption{{\it AliGenerator} is the base class which has the responsibility
 to generate the primary particles of an event. Some realisations of
 this class do not generate the particles themselves but delegate the
 task to an external generator like PYTHIA through the {\it TGenerator}
 interface.}
\label{fig:aligen}
\end{figure}

\subsection{Interfaces}

To facilitate the usage of different generators we have developed an
abstract generator interface called {\it AliGenerator}, see
Fig.~\ref{fig:aligen}.  
The objective is to provide the user with an
easy and coherent way to study a variety of physics signals as well as
full set of tools for testing and background studies. This interface
allows the study of full events, event-by-event analysis, single
processes and a mixture of both, {\it Cocktail Events}.

During creation of a  {\it AliGenerator} object it registers itself to the run steering 
object.  
The latter owns and provides access to the particle stack of type {\it AliStack}, 
so that the method {\it AliGenerator::Generate()} can write particles to it.   
Alternatively a stack can be directly connected to the {\it AliGenerator} object in order 
to use particle production stand-alone for fast simulation tasks or in order to create several events 
before a combination of them is written to the main stack.

\subsubsection{External generators}

Several event generators are available via the abstract ROOT class
that implements the generic generator interface, {\it TGenerator}.  
Through implementations of this abstract base class we wrap FORTRAN MonteCarlo
codes like PYTYIA, HERWIG and HIJING that are thus accessible from the
AliRoot classes. 
In particular the interface to PYTHIA used with
PDFLIB includes the use of nuclear structure functions.

{\it AliGenerator} derived classes like {\it AliGenPythia} or {\it AliGenHijing} combine the external 
particle production by delegation to {\it TGenerator} with their main task writing particles to the stack.
In addition simplified configuration of the external generators is provided. Predefined simulation 
configurations like minimum bias or jet-production help the user to perform easily simple simulation 
tasks. 
 
An interesting byproduct of this design is that in a ROOT session the external generators can be used interactively. 
Run time configuration, particle production and analysis can be performed from the command-line interface.

\begin{figure}[htb]
\centering
\includegraphics[width=65mm]{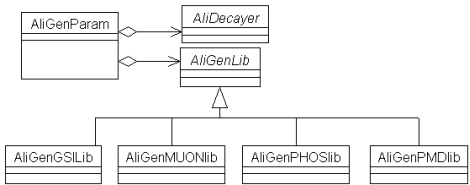}
\caption{{\it AliGenParam} is a realisation of {\it AliGenerator} that generates
particles using parameterised transverse momentum and pseudo rapidity
distributions. Instead of coding a fixed number of parameterisations
directly into the class implementations, user defined parametrisation
libraries ({\it AliGenLib}) can be connected at run time allowing for
maximum flexibility.}
\label{fig:evglib}
\end{figure}

\begin{figure}[htb]
\centering
\includegraphics[width=65mm]{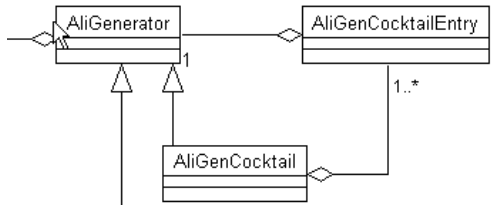}
\caption{The {\it AliCocktail} generator is a realisation of {\it AliGenerator}
which does not generate particles itself but delegates this task to a
list of objects of type AliGenerator that can be connected as entries
(AliGenCocktailEntry) at run time. In this way different physics
channels can be combined in one event.}
\label{fig:cocktail}
\end{figure}

\subsubsection{Parameterisations}

In many cases, the expected transverse momentum and rapidity
distributions of particles are known. In other cases the effect of
variations in these distributions must be investigated. 
In both situations it is appropriate to use generators that produce
primary particles and their decays sampling from parametrised spectra.
To meet the different physics requirements in a modular way, the
parameterisations are stored in independent function libraries wrapped
into classes that can be plugged into the generator. This is
schematically illustrated in Fig.~\ref{fig:evglib} where four
different generator libraries can be loaded via the abstract generator
interface.

\subsubsection{Event cocktails}

It is customary in heavy ion event generation to superimpose different
signals on an event to tune the reconstruction algorithms. 
In other cases the user needs the possibility to assemble events by combining 
different particle types and parameterisations.

This is possible in AliRoot via the so-called {\it cocktail generator} (Fig.~\ref{fig:cocktail}).
This creates events from user-defined particle cocktails by choosing as ``ingredients'' a list of 
generators.
Through its recursive design {\it AliGenCocktail} is at the same time a generator and 
a container class for generators. The generation task is performed by delegating 
to the registered generators.

Another important application of {\it AliGenCocktail} has been found for the simulation of 
p-A collisions. Generators like HIJING can simulate p-A collisions but they do not simulate 
slow nucleons. The detection of these slow nucleons is important for the determination of the 
centrality of the collisions (event geometry). Simple parameterisations exist for their production we 
have designed base classes {\it AliGenSlowNucleons} and {\it AliSlowNucleonModel} for this purpose.
They need, however, the event collision geometry as an input. 
An easy solution was to combine the instance of the 
slow nucleon generator with a generator that can provide a collision geometry using {\it AliGenCocktail}.
Communication between the two is via the class {\it AliCollisionGeometry}.

\subsubsection{Afterburner}

The modularity of the event generator implementation allows at the same time easy
integration with the simulation steering class {\it AliRun} and with the
objects that are responsible for changing the output of event
generators or for assembling new events making use of the input of
several events. 
These processors are generally called {\it Afterburners}.  
They are especially needed to introduce a controlled
(parameterised) particle correlation into an otherwise uncorrelated
particle sample. In AliRoot this task is further simplified by the
implementation of the stack class {\it AliStack} that can be connected to
both {\it AliRun} and {\it AliGenerator}. 
Currently afterburners are used for two particle correlation simulations 
and analysis and for the simulation of azimuthal flow.

\section{Fast simulation}

Detailed detector simulation is needed together with reconstruction to
evaluate the acceptance, efficiency and resolution for specific
physics probes.  
Owing to the high particle multiplicity per event
this analysis can be based on a few thousand events. 
Many types of physics analysis, such as D meson reconstruction from hadronic decay
channels, and trigger studies have to make use of millions of events.
Computing resources are in general not available for such high statistics simulations.

To reach the required sample size, fast simulation methods based on
meaningful parameterisations of the results from detailed and
consequently slow simulations are applied. The systematic error
introduced by the parameterisations is in general small compared to
the reduction of the statistical error. This is particularly true for
the studies of the invariant mass continuum below a resonance.

It is hard to find a common denominator for fast simulation methods
since they are very specific to the analysis task. As a minimum
abstraction, we have designed base classes that allow for a
representation of the detector or detector systems as a set of
parameterisations of acceptance, efficiency and resolution. The Muon
Spectrometer fast simulation has been implemented using these classes.

Another interesting development concerns the fast simulation of the
resolution and efficiency of tracking in the central barrel. In this
approach full tracking is performed for the inner tracking system
(ITS), which is needed for detailed secondary vertex reconstruction
studies. After this, TPC resolution and efficiency are obtained from
the track parameters at the inner radius of the TPC, using a
parameterisation.

\section{Conclusions}

The ALICE Offline Project has developed a coherent simulation framework for 
detector and physics performance studies in the AliRoot Framework based on ROOT. 
The main simulation components are the Virtual MC Interface and 
event generator interfaces tailored to the needs of the heavy ion community

The framework has proven its maturity and functionality in the large MC production that has been performed
for the studies related to the ALICE Physics Performance Report.
The framework is still confronted with new requirements from both physics studies 
and new detector elements under study, most recently the proposed electromagnetic calorimeter
together with the ideas of studying  jet physics with the ALICE detector. 
The easiness with which such new elements can be introduced are the consequence of our modular approach.


\begin{thebibliography}{9}

\bibitem{aliroot}
R.~Brun, P.~Buncic, F.~Carminati, A.~Morsch, F.~Rademakers, K.~Safarik on behalf of the ALICE collaboration, 
``The AliRoot framework, status and perspectives'', in these proceedings.

\bibitem{root}
  http://root.cern.ch 

\bibitem{geant3}
  R.~Brun, F.~Bruyant, M.~Maire, A.C.~McPherson, P.~Zanarini, GEANT3
  User Guide, CERN Data Handling Division DD/EE/84-1 (1985) 

\bibitem{fluka}
    A.Fass\`o, A.Ferrari, P.R.Sala, {\it"Electron-photon transport in
      FLUKA: status"}, Proceedings of the MonteCarlo 2000 Conference,
      Lisbon, October 23-26 2000, A.Kling, F.Barao, M.Nakagawa,
      L.Tavora, P.Vaz - eds., Springer-Verlag Berlin, p.159-164 (2001).
\\
    A.Fass\`o, A.Ferrari, J.Ranft, P.R.Sala, {\it"FLUKA: Status and
      Prospective for Hadronic Applications"}, Proceedings of
      the MonteCarlo 2000 Conference, Lisbon, October 23-26 2000,
      A.Kling, F.Barao, M.Nakagawa, L.Tavora, P.Vaz - eds. ,
      Springer-Verlag Berlin, p.955-960 (2001).

\bibitem{geant4}
S. Agostinelli et al., "Geant4 - A Simulation Toolkit", CERN-IT-20020003, KEK Preprint 2002-85, SLAC-PUB-9350, submitted to Nuclear
Instruments and Methods A.


\bibitem{vmc} D.~Adamova, V.~Berejnoi, R.~Brun, F.~Carminati, A.~Fass\`o, E.~Fut\'o, I.~Gonzalez, I.~Hrivnacova, A.~Morsch 
on behalf of the ALICE Offline Project, ``The Virtual MonteCarlo'', in these proceedings.

\bibitem{tgeo} R.~Brun, A.~Gheata, and  M.~Gheata on behalf of the ALICE Offline Project, ``A geometrical modeler for HEP'', 
in these proceedings.

\bibitem{flugg}   M.~Campanella, A.~Ferrari, P.R.~Sala, and S. Vanini,
  ``Reusing Code from FLUKA and GEANT4 geometry",
  ATLAS Internal Note ATL-SOFT 98-039 (1998) \\
  M.~Campanella, A.~Ferrari, P.R.~Sala and S.~Vanini, ``First
  Calorimeter Simulation with the FLUGG prototype'',
  ATLAS Internal Note ATL-SOFT-99-004 (1999)

\end{thebibliography}
\end{document}